\documentclass{moriond}

\def\be{\begin{equation}}
\def\ee{\end{equation}}
\def\bea{\begin{eqnarray}}
\def\eea{\end{eqnarray}}

\def\geff{G_{\rm eff}}

\def\gn{G_{\rm N}}

\def\gsp{\eta}
 
\def\fs{f\sigma_8} 
\def\omo{\Omega_m^0}
\def\vv{\vskip2mm}

\usepackage{amsmath, amsfonts, amssymb, amsthm} 

\newcommand{\Photo}{}

\begin{document}
\vspace*{4cm}
\title{GENERAL FEATURES OF SINGLE-SCALAR FIELD \\ DARK ENERGY MODELS \footnote{Proceedings of 51st Rencontres de Moriond}}

\author{ L. PERENON }

\address{Aix-Marseille Universit\'e, CNRS, CPT, UMR 7332, 13288 Marseille,  France. \\
Universit\'e de Toulon, CNRS, CPT, UMR 7332, 83957 La Garde,  France.}

\maketitle
\abstracts{We present a systematic study of modified gravity (MG) models containing a single scalar field non-minimally coupled to the metric. Despite a large parameter space, exploiting the effective field theory of dark energy (EFT of DE) formulation and imposing simple physical constraints such as stability conditions and (sub-)luminal propagation of perturbations, we arrive at a number of generic predictions about the large scale structures.}

\begin{figure}[!]
	\begin{center}\vskip-2mm
	\includegraphics[scale=0.42]{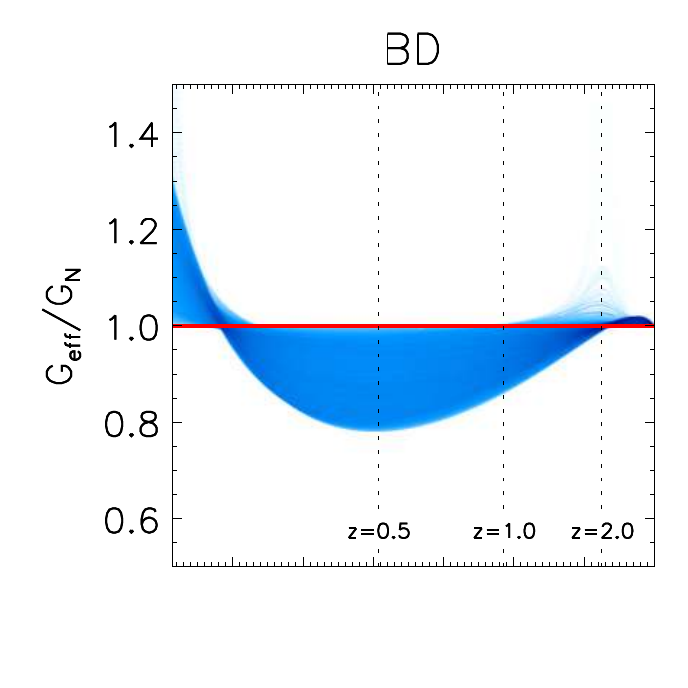} 	  \hskip-4mm
	\includegraphics[scale=0.42]{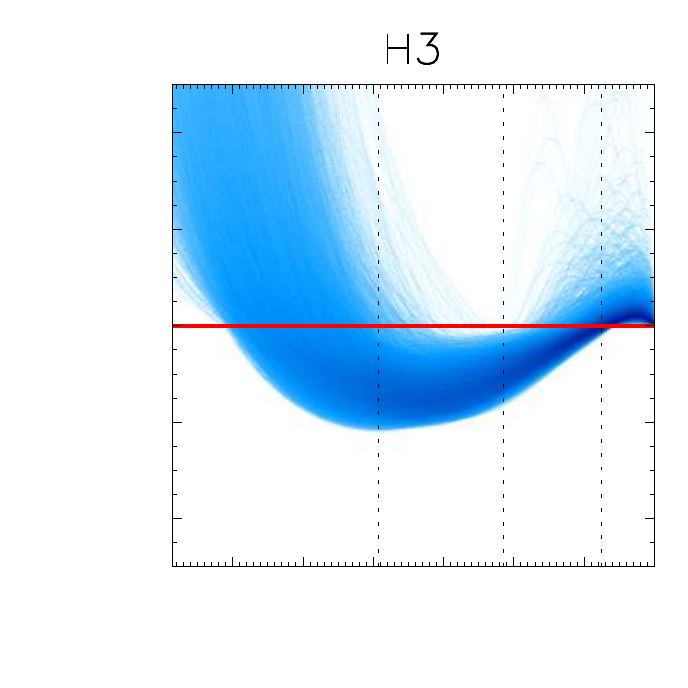}    \hskip-4mm
	\includegraphics[scale=0.42]{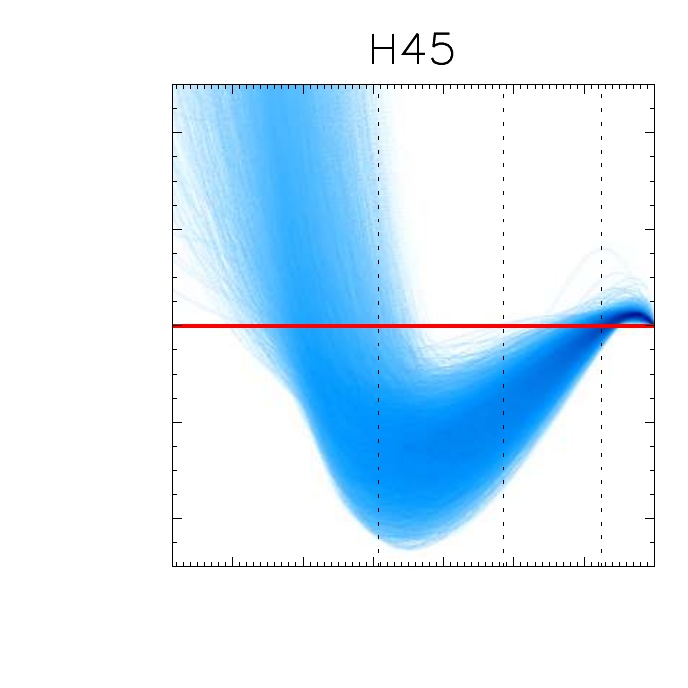}   \vskip-8mm
	\includegraphics[scale=0.42]{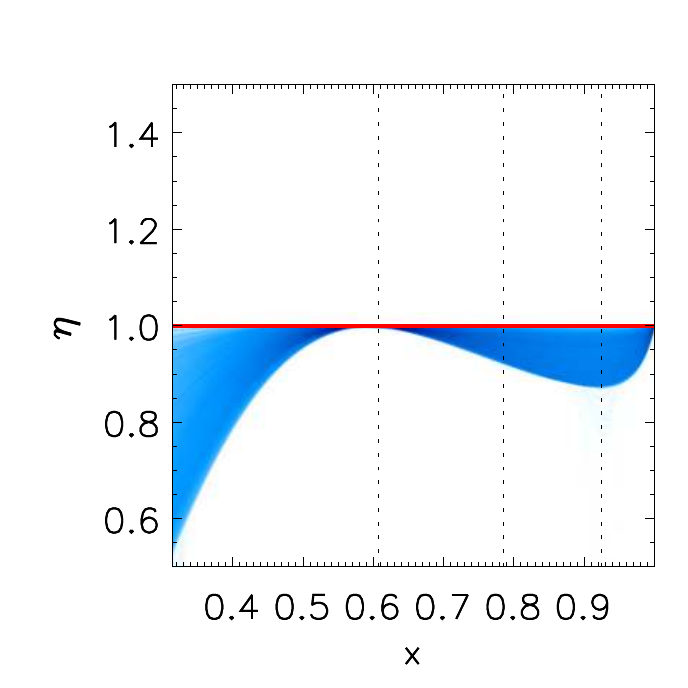}       \hskip-4mm
	\includegraphics[scale=0.42]{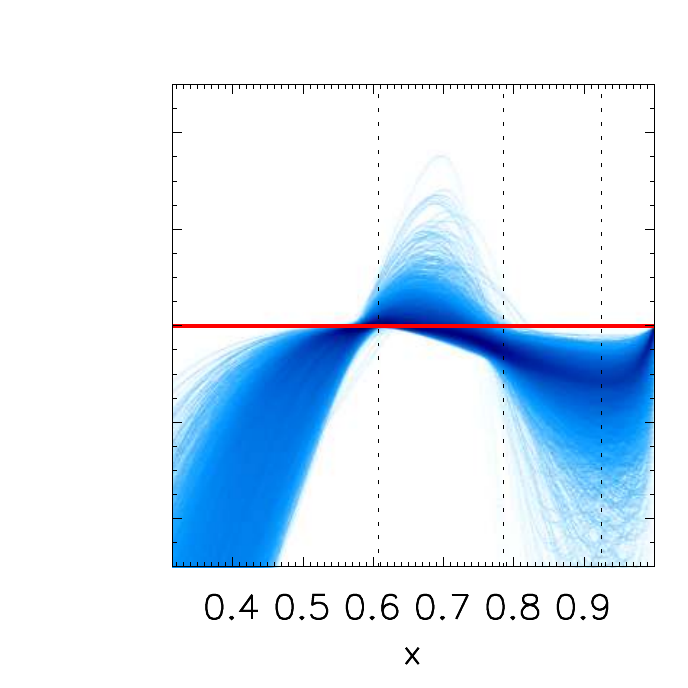}       \hskip-4mm
	\includegraphics[scale=0.42]{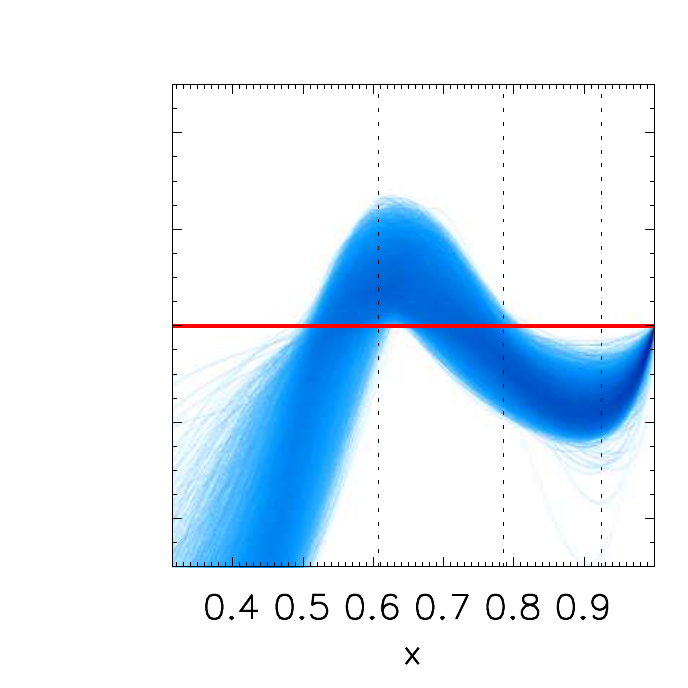}      \vskip-4mm
	\includegraphics[scale=0.42]{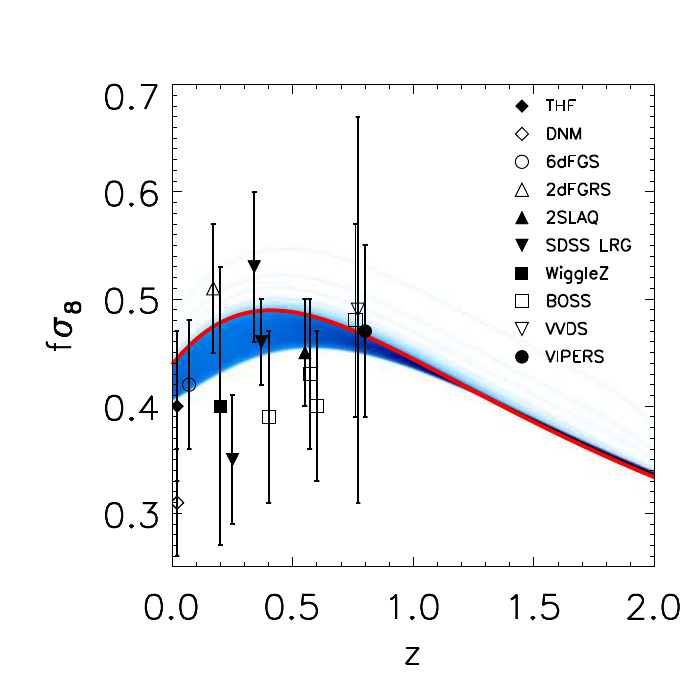}     \hskip-4mm
	\includegraphics[scale=0.42]{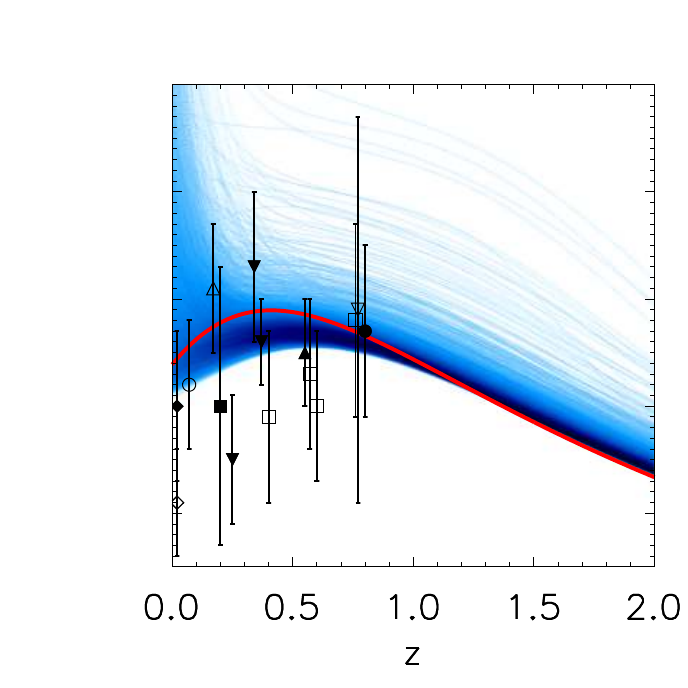}     \hskip-4mm
	\includegraphics[scale=0.42]{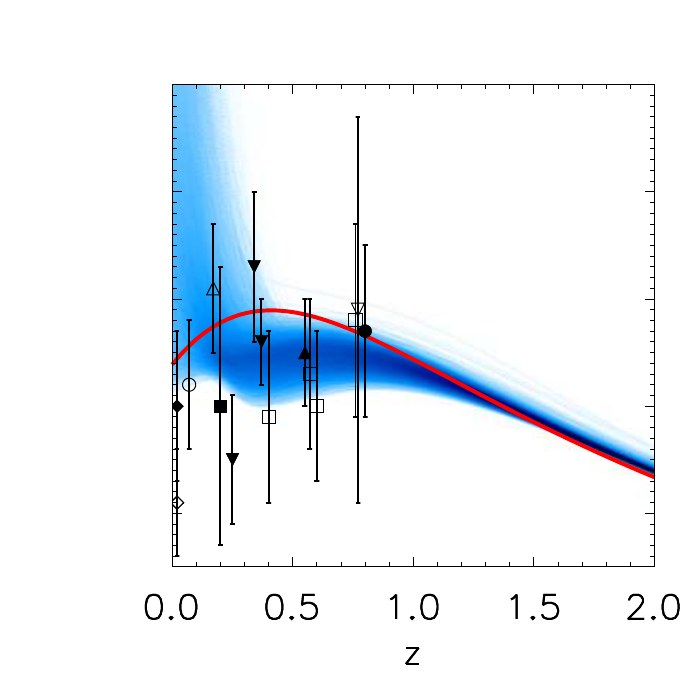}\vskip-2mm
\caption{The behaviour of $\geff/\gn$ (first row), $\gsp$ (second row) as a function of the reduced matter density $x$ and the redshift evolution of $\fs$ (third row) are shown for $10^4$ randomly generated viable EFT models (\emph{BD}, \emph{H3}, \emph{H45}). The dotted vertical lines identify, from left to right, the cosmic epochs $z=0.5$, $z=1$ and $z=2$. The thick red line represents the $\Lambda$CDM prediction. The density of curves passing through a region is shown by the levels of blue.}\vskip-2mm\vskip-2mm 
	\label{fig}   
	\end{center}
\end{figure}

The goal of this work, in collaboration with F. Piazza, C. Marinoni and L. Hui \cite{Perenon:2015sla}, is to study the predictability of MG theories aiming at challenging the standard $\Lambda$CDM explanation of cosmic acceleration. We use the EFT of DE \cite{EFTOr} for it has established a common formalism to describe the widest set of MG theories, those adding a single extra scalar degree of freedom to the Einstein-Hilbert action. 
\vv
Such a unifying description enables MG theories to be parametrized in a common framework in terms of structural functions of time, describing how matter perturbations evolve in the universe. The requirements needed to fully describe an EFT model can be reduced to two constants and three functions of time $
\left\{\omo,\; w, \; \mu(t), \; \mu_3(t),\; \epsilon_4(t) \right\}$. The three functions are non-minimal couplings, once ``turned on" they enable the description theories in the Horndeski class: $\mu$ is the Brans-Dicke (\emph{BD}) non-minimal coupling, adding $\mu_3$ models the cubic galileon (\emph{H3}) term and $\epsilon_4$ encodes the 4 and 5 Horndeski (\emph{H45}) Lagrangians. In parallel, the EFT of DE allows one to independently set the background expansion, $i.e$ the Hubble rate. This reduces to fixing the two constants, the present fractional matter density of non-relativistic matter in the perfect fluid approximation $\omo$ and the background DE equation of state parameter $w$ \cite{pheno}. They are set by the latest constraints \cite{planck} to reproduce a flat $\Lambda$CDM background, thus respectively $\sim 0.3$ and $-1$. 
\vv
Another asset of the EFT of DE is to provide a clear and common means of assessing if theories are pathological or not, $i.e$ whether they suffer from ghost or gradient instabilities. The main purposes of this work is to show that despite large degrees of freedom, definite features common to all healthy ---stable and with no superluminal propagation of scalar and tensor modes--- EFT models arise within the vast Horndeski class. We show this by expanding the non-minimal coupling functions in power series up to second order in the reduced matter density, $ x $, used as time variable. An overall $(1-x)$ pre-factor ensures the vanishing of the couplings at early times, hence recovering general relativity. We randomly generate the coefficients of the expansions until we obtain $10^4$ healthy \emph{BD}, \emph{H3} and \emph{H45} theories.
\vv
At the linear level and under certain conditions, non-standard gravitational scenarios result in time dependent modifications of the Newton's constant $\geff$ and of the gravitational slip parameter $\gsp$. The requirement of a healthy theory leads to bounded evolutions and generic features in these quantities. For instance, as is it shown in fig.~\ref{fig}, the curves of $\geff/\gn$ display a characteristic {\it S-shape}, an alternation, stronger gravity at early times, weaker gravity at intermediate redshifs and stronger gravity again today. Two competing effects are at play: the matter density entering the perturbations of an EFT model is not the same as that of the background $x$ and being generally smaller it lowers the gravitational coupling; the contribution of the extra scalar field is attractive (healthy spin-0 field) which enhances the gravitational coupling. 

The effective Newton constant acts as a source term in the growth of matter perturbations. An observable of the large scale structures such as the growth function $\fs$ can be effectively used to constrain MG. As it is shown in fig.~\ref{fig}, its sensitivity to $\geff$ translates into having almost all models displaying lower growth than $\Lambda$CDM at $0.5\lesssim z\lesssim 1$ and always predicting stronger growth for $z>1.5$. The \emph{H3} and \emph{H45} cases, exhibiting more freedom, can introduce a few models deviating strongly from the $\Lambda$CDM reference however they are statistically insignificant.

Complementing this analysis with $\gsp$ highlights more definite features and enhances the discriminating power. For \emph{BD} theories, one can show analytically that $\gsp \leqslant 1$ always holds. When going to \emph{H3} and \emph{H45}, a distinctive shape is still seen, $\eta$ is always smaller than unity at any redshifts except in the window $0.5\lesssim z\lesssim 1$.
\vv
We are now exploring, Perenon $et$ $al.$ in prep., the impact on the universality of our results when changing the background DE equation of state and changing the asymptotic behaviour of the non-minimal couplings (early dark energy). We are also investigating the link between self-acceleration and the production of weaker gravity.

\vskip-4mm\vskip-4mm

\section*{References}\vskip-3mm\vskip-1mm
\begin{thebibliography}{99}
\bibitem{Perenon:2015sla} 
  L.~Perenon, F.~Piazza, C.~Marinoni and L.~Hui,
  JCAP {\bf 1511}, no. 11, 029 (2015)
  doi:10.1088/1475-7516/2015/11/029
  [arXiv:1506.03047 [astro-ph.CO]].
\bibitem{EFTOr} 
  G.~Gubitosi,  F.~Piazza and F.~Vernizzi,
  JCAP {\bf 1302}, 032 (2013)
  [arXiv:1210.0201 [hep-th]].     
\bibitem{pheno} 
  F.~Piazza, H.~Steigerwald and C.~Marinoni,
  JCAP {\bf 1405}, 043 (2014)
  [arXiv:1312.6111 [astro-ph.CO]].
\bibitem{planck} 
  P.~A.~R.~Ade {\it et al.}  [Planck Collaboration],
  [arXiv:1502.01589 [astro-ph.CO]].
\end{thebibliography}
\end{document}